\documentclass[preprint,12pt]{elsarticle}

\newcommand{\G}{\mathcal{G}}
\newcommand{\V}{\mathcal{V}}
\newcommand{\E}{\mathcal{E}}

\newcommand{\R}{\mathcal{R}}
\newcommand{\N}{\mathcal{N}}

\usepackage{graphicx}
\usepackage{amsmath}
\usepackage{float}
\usepackage{color}
\usepackage{graphicx}
\usepackage{pdfpages}
\usepackage{epstopdf}
\usepackage{subfig}
\newtheorem{remark}{Remark}

\graphicspath{ {images/} }

\begin{document}

\begin{frontmatter}



\title{Scheduling in Wireless Networks with Spatial Reuse of Spectrum as Restless Bandits}


\author[VB]{Vivek S. Borkar} 
\author[SC]{Shantanu Choudhary}
\author[VB]{Vaibhav Kumar Gupta}
\author[VB]{Gaurav S. Kasbekar\fnref{3}}

\address[VB]{Department of Electrical Engineering, Indian Institute of Technology (IIT), Bombay.}
\address[SC]{Hyundai Mobis, Hyderabad.}
\fntext[3]{Their email
addresses are \{borkar, vaibhavgupta, gskasbekar\}@ee.iitb.ac.in, choudharyshantanu92@gmail.com resp. VB was supported by a J.\ C.\ Bose Fellowship. SC
worked on this research while he was with IIT Bombay.}



\begin{abstract}
We study the problem of scheduling packet transmissions with the aim of minimizing the energy consumption and data transmission delay of users in a wireless network in which spatial reuse of spectrum is employed. We approach this problem using the theory of Whittle index for cost minimizing restless bandits, which has been used to effectively solve problems in a variety of applications. We design two Whittle index based policies-- the first by treating the graph representing the network as a clique and the second based on interference constraints derived from the original graph. We evaluate the performance of these two policies via extensive simulations, in terms of average cost and packets dropped, and show that they outperform
the well-known Slotted ALOHA,  maximum-weight scheduling and Lyapunov drift algorithms.
\end{abstract}


\end{frontmatter}

\section{INTRODUCTION}
Several wireless networks employ \emph{spatial reuse} of spectrum, i.e., in such networks,  multiple mutually far apart transmitters simultaneously  send data to their respective receivers on the same channel without interfering with each other. Examples of such networks  are wireless cellular networks, mesh networks, ad hoc and sensor networks. They have a variety of applications, e.g., in military and emergency operations, Internet access for communities, intrusion detection, precision agriculture, environmental monitoring and industrial monitoring~\cite{RF:mesh:networks:survey:akyildiz,RF:WSN:survey:akyildiz,RF:Atzori:IoT:Survey,RF:ad:hoc:perkins}.
Since nodes in such networks are often battery-powered, a key objective is to {\em achieve energy efficiency}~\cite{RF:WSN:survey:akyildiz,RF:Atzori:IoT:Survey}. Another important objective is to {\em minimize the data transmission delay}, especially that of real-time traffic such as audio and video calls, emergency alerts from security systems etc. Also, a basic function in wireless networks that employ spatial reuse of spectrum is \emph{scheduling}, i.e., selecting a mutually non-interfering set (independent set)~\cite{RF:graph:theory:west} of nodes that will transmit in each time slot~\cite{RF:sharma:shroff:complexity:scheduling}. In this paper, we address the fundamental problem of scheduling packet transmissions with the objective of minimizing the energy consumption and data transmission delay of users in wireless networks in which spatial reuse of spectrum is employed.

Scheduling in wireless networks that employ spatial reuse of spectrum has been extensively studied in the research literature-- see~\cite{RF:neely:survey,c4,RF:berry:energy:efficient:survey} for  surveys. A throughput-optimal scheduling policy was provided in the seminal work~\cite{RF:tassiulas:stability}. The complexity of throughput-optimal scheduling in multi-hop wireless networks subject to inteference constraints was studied in~\cite{RF:sharma:shroff:complexity:scheduling}. In~\cite{RF:chaporkar:maximal:scheduling}, it was shown that a distributed scheduling strategy, called maximal scheduling, attains a guaranteed fraction of the maximum throughput region in multi-hop wireless networks. In~\cite{RF:gupta:low:complexity}, distributed scheduling schemes were designed  that achieve throughput close to that of maximal schedules, but whose complexity is low. A distributed scheduling scheme that guarantees maximum throughput in multi-hop wireless networks was presented in~\cite{RF:Modiano:gossiping}. However, the schemes in~\cite{RF:sharma:shroff:complexity:scheduling,RF:tassiulas:stability,RF:chaporkar:maximal:scheduling,RF:gupta:low:complexity,RF:Modiano:gossiping} were designed so as to maximize the achievable throughput. In contrast, in this paper our objective is to design a scheduling scheme that minimizes the delay and energy consumption. A large number of medium access control (MAC) protocols, including the well-known Pure ALOHA, Slotted ALOHA, CSMA/ CA and IEEE 802.11 Distributed Coordination Function protocols, have been designed for wireless networks-- see~\cite{RF:kumar:mac:protocol:survey} for a survey. These MAC protocols can be used for scheduling in a  wireless network that employs spatial reuse of spectrum. These protocols, however, do not in general minimize the energy consumption or delay.

Scheduling in wireless networks with the objectives of minimizing the energy consumption and/ or delay has been extensively studied in prior work. A survey of schemes for delay-aware resource control in a multi-hop wireless network is provided in~\cite{RF:cui:delay:aware:survey}. A scheduling scheme for minimizing the energy-expenditure in a time-varying wireless network with adaptive transmission rates has been provided in~\cite{RF:energy:optimal:neely}. In~\cite{RF:neely:energy:delay:tradeoffs}, the problem of allocating power to links as a function of current channel states and queue backlogs to stabilize the system while minimizing energy expenditure and maintaining low delay in a multiuser network is studied.  In~\cite{RF:sadiq:delay:optimal}, the problem of designing opportunistic scheduling policies that minimize the average delay in a wireless network with multiple users sharing a wireless channel is studied. In~\cite{RF:salodkar:multiuser:scheduling}, energy-efficient scheduling with delay constraints in a multiuser wireless network is studied. The problem of delay minimization under power constraints for uplink transmission in a multiuser wireless network is studied in~\cite{RF:moghadari:delay:optimal}.
The problem of minimizing the transmission power subject to
a delay constraint in a multiuser wireless network is studied in~\cite{RF:zhang:power:delay:tradeoff}. However, with the exception of our prior work~\cite{c1},  no work has addressed the problem of scheduling in a wireless network with the objective of minimizing the energy consumption and delay using the theory of \emph{Whittle index}~\cite{Whittle}. In \cite{c1}, at most one user can successfully transmit at a time on the channel. In this paper, we study a wireless network that employs spatial reuse of spectrum, allowing multiple simultaneous transmissions.

Specifically, we consider a  wireless network with multiple users deployed in a region and communicating using a single channel. We represent the network using an undirected graph~\cite{RF:graph:theory:west}, in which there is a node representing each user, and there is an edge between two nodes iff the transmissions of the corresponding users interfere. In each time slot,   an independent set of users which will transmit in the slot needs to be selected. The energy consumed when a user transmits on the channel is modeled by an \emph{``energy cost''}, which is an increasing function of the number of packets transmitted. Note that the delay experienced by a packet is an increasing function of the number of packets ahead of it in its queue. Since we seek to minimize packet delays, we also consider a cost proportional to the queue length, referred to as the \emph{``holding cost''}. The cost incurred in a slot at a user is the sum of the energy cost and the holding cost.  Our objective is to minimize the time-averaged total cost incurred at all the users in the network. To solve this problem, we use the \emph{Whittle index policy}, which was introduced in~\cite{Whittle} and has been used to effectively solve problems in a variety of applications~\cite{ephemeral, Cloud, proc_sharing, Cowan, Gittins, Jacko, Larr,  Liu, Nino, Ny, Raghu, Ruiz}. Specifically, the constraint that an independent set of users must transmit in each time slot makes the above cost minimization problem provably hard~\cite{Papa}. So we relax this constraint to a time-averaged constraint and formulate a corresponding unconstrained problem using Lagrange multipliers. Using a technique similar to that introduced by Whittle~\cite{Whittle}, we decouple this unconstrained problem into individual problems for each user and define suitable Whittle-like indices. In particular, we design two Whittle index based policies-- the first by treating the graph representing the network as a clique (i.e., a complete graph) and the second based on interference constraints derived from the original graph. Also, we propose a distributed algorithm for activating users, which can be used to select an independent set of users to activate in a time slot after Whittle indices of all the users have been computed. We evaluate the performance of the above two policies via extensive simulations, in terms of average cost and packets dropped, and show that they outperform the well-known \emph{Slotted ALOHA}~\cite{c7}, \emph{maximum-weight scheduling (MWS)}~\cite{RF:tassiulas:stability} and \emph{Lyapunov drift}~\cite{RF:energy:optimal:neely} algorithms.

The rest of this paper is organized as follows. In Section~\ref{SC:model:prob:formulation:background}, we describe the model and problem formulation, briefly review the theory of Whittle index, and explain the differences between the contributions of this paper and our prior work~\cite{c1}. We present two scheduling algorithms based on Whittle indices to solve the above problem in Section~\ref{SC:algorithms}. We present simulation results in Section~\ref{SC:simulations} and provide conclusions and directions for future research in Section~\ref{SC:conclusions}.

\section{Model, Problem Formulation and Background}
\label{SC:model:prob:formulation:background}
\subsection{Model and Problem Formulation}
\label{SSC:model:prob:formulation}
\begin{figure}[h]%
    \centering
    \subfloat{\label{1b}{\includegraphics[width=4cm, height=3.7cm]{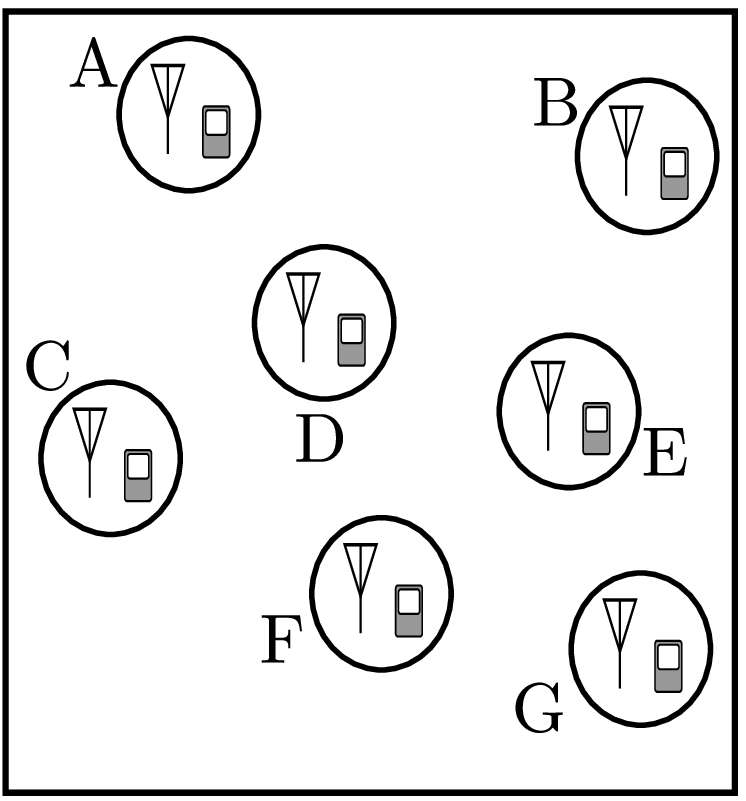} }}%
    \hspace{.1cm}
    \subfloat{\label{1c}{\includegraphics[width=4cm, height=3.7cm]{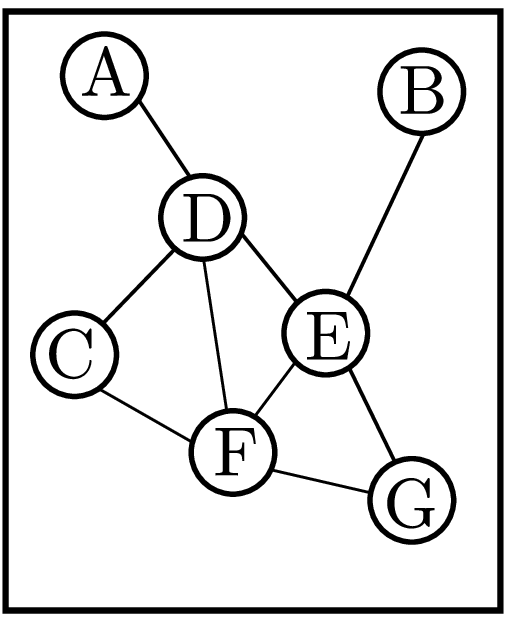} }}%
    \caption{The figure on the left shows a wireless network with $L = 7$ users (transmitter--receiver pairs) and the figure on the right shows the undirected graph used to represent the network.}%
    \label{model}%
\end{figure}

We consider a wireless network consisting of $L$ users deployed in a region and communicating using a single channel. Each user is a transmitter-receiver pair, with a queue at the transmitter of packets which need to be sent to the receiver. Recall that the wireless medium has the property that simultaneous transmissions by two users that are close to each other interfere with each other, whereas the channel can be simultaneously used at mutually far-apart locations without interference. To model these spatial reuse (interference) constraints, we represent the network using an undirected graph $\G = (\V, \E)$, in which $\V$  is the set of users and there is an edge between two users $i, j \in \V$ iff the transmissions of users $i$ and $j$ interfere with each other (see Fig.~\ref{model}). Let $\N(i)$ be the set of neighbors of user $i$, i.e., the set $\{j \in \V : \exists \ (i,j) \in \E\}$.

Time is divided into slots of equal durations. The queue of user $i$ evolves according to the dynamics:
\begin{equation}
X^i_{n+1} = \left[X^i_n - \nu^i_n (X^i_n\wedge\Psi^i) + \xi_{n+1}^i\right]\wedge M^i, \label{dynamics}
\end{equation}
where $X^i_n$ is the length of the queue of user $i$ in time slot $n$, $\xi_n^i$ is the number of arrivals at the queue of user $i$ in time slot $n$, $M^i$ is the capacity of the buffer of user $i$, $\Psi^i$ is the maximum number of packets that may be transmitted by user $i$ in a slot and $\nu^i_n$ is $1$ if user $i$ transmits in slot $n$ and $0$ otherwise. We say that a user is ``active'' in a slot if it transmits and ``passive'' if not. We assume that the number of packet arrivals, $\xi_n^i$, $n = 0, 1, 2, \ldots$, in different slots are independent and identically distributed  (IID) random variables with distribution  $\mu^i(\cdot)$.

The cost of holding packets in the queue of user $i$ is $C^i$ per packet per slot. That is, if there are $x$ packets in queue  $i$ in a given slot, then a cost of $x C^i$ is incurred.  This cost models the delay requirement of a queue. In particular, the higher the value of $C^i$, the more stringent the delay requirements of the packets stored in queue $i$. For example, the value of $C^i$ may be set to a low (respectively, high) value if queue $i$ stores elastic traffic such as file transfer packets (respectively, real-time traffic such as audio and video flow packets). Let $f^i(z)$ be the ``energy cost'', i.e., the cost incurred by user $i$ due to expenditure of energy  when it transmits $z$ packets.

Let $\N^*(i) := \N(i)\cup\{i\}$. If two or more users from the set $\N^*(i)$ transmit in time slot $n$, their transmissions may interfere with each other, leading to the constraints:
\begin{equation}
\label{EQ:interference:constraints}
\sum_{j \in \N^*(i)}\nu^j_n \leq 1, \ \forall i.
\end{equation}
If a subset of the users in $\V$ transmits in a time slot subject to \eqref{EQ:interference:constraints}, then that subset constitutes an \emph{independent set}\footnote{Recall that an \emph{independent set}~\cite{RF:graph:theory:west}  in a graph is a set of nodes such that there is no edge between any pair of nodes in the set.}
of nodes in the graph $\G = (\V, \E)$.
Let $Z_n^i := X^i_n\wedge\Psi^i$. We seek to minimize the time-averaged cost incurred by all  users, i.e.:
\begin{equation}
\label{EQ:objective:function}
\min \lim_{N \uparrow \infty}\frac{1}{N} \sum_{n=0}^{N-1} \sum_{i \in \V} E \left[ \nu_n^i f^i(Z_n^i) + C^i X_n^i \right],
\end{equation}
subject to the interference constraints \eqref{EQ:interference:constraints}. That is, our objective is to select an independent set of users to activate, subject to \eqref{EQ:interference:constraints}, in each time slot so as to minimize  \eqref{EQ:objective:function}.

\begin{remark}
The constraint in \eqref{EQ:interference:constraints} may prevent two mutually non-interfering users from simultaneously transmitting. For example, consider the graph with $\V = \{1,2,3\}$ and $\E = \{(1,2),(1,3)\}$. The constraint in \eqref{EQ:interference:constraints} for user $i = 1$ is:
\[
\nu_n^1 + \nu_n^2 + \nu_n^3 \leq 1.
\]
This constraint is violated if users $2$ and $3$ simultaneously transmit in slot $n$. However, note that users $2$ and $3$ are mutually non-interfering.

Nevertheless, to facilitate the following analysis, we impose the constraint  \eqref{EQ:interference:constraints}. In Section~\ref{SSC:activation:algorithm}, we provide an algorithm for activating users in different time slots that ensures that each user from a \emph{maximal} independent set of users transmits in every time slot.
\end{remark}

\subsection{Background on Whittle index}
\label{SSC:whittle:index:background}
We briefly recall here the basics of Whittle index~\cite{Whittle} for cost minimizing restless bandits. The latter refers to a collection of $N \geq 2$ controlled Markov chains $Y^i_n, n \geq 0$, $i \in \{1, \ldots, N\}$, taking values in discrete state spaces $S^i$, with two modes of operation, active and passive, with corresponding transition probabilities given by $p_{i,1}(t|s), \ p_{i,0}(t|s)$ and running costs $c_1(s), c_0(s)$ resp., where $s,t \in S^i$. The control process associated with $i$th chain is $u_i(n), n \geq 0,$ taking values in $\{0,1\}$ with the interpretation that value $1$ (resp., $0$) corresponds to active (resp., passive) mode. Thus the transition probability at time $n$ for the $i$th process is $p_{i,u_i(n)}( \cdot | Y^i_n)$. The objective is to minimize the average cost
$$\limsup_{n\uparrow\infty}\frac{1}{n}E\left[\sum_{m=0}^{n-1}\sum_ic_{u_i(m)}(Y^i_m)\right]$$
subject to the per stage constraint
$$\sum_iu_i(n) \leq M \ \forall n$$
for some $1 < M < N$, which couples the problems. This constraint makes the problem provably hard~\cite{Papa}. The Whittle device~\cite{Whittle} is to relax it to the average constraint
$$\limsup_{n\uparrow\infty}\frac{1}{n}E\left[\sum_{m=0}^{n-1}\sum_iu_i(m)\right] \leq M$$
and consider the unconstrained problem of minimizing
\begin{equation}
\limsup_{n\uparrow\infty}\frac{1}{n}E\left[\sum_{m=0}^{n-1}\sum_i(c_{u_i(m)}(Y^i_m) + \lambda u_i(m))\right], \label{together}
\end{equation}
where $\lambda$ is the Lagrange multiplier. Given $\lambda$, this decouples into individual control problems of minimizing
\begin{equation}
\limsup_{n\uparrow\infty}\frac{1}{n}E\left[\sum_{m=0}^{n-1}(c_{u_i(m)}(Y^i_m) + \lambda u_i(m))\right] \label{separate}
\end{equation}
for each $i$. Whittle uses this to motivate the so called Whittle index as follows. The problem is said to be (Whittle) indexable if the set of passive states (i.e., the states $Y^i_m$ for which $u_i(m) = 0$ is the optimal action) for each individual problem $i$ monotonically decreases from the whole state space to the empty set as the `tax' $\lambda$ decreases from $+\infty$ to $-\infty$. If so, the Whittle index for the $i$th problem is the function $\lambda^i : S^i \mapsto \R$ such that $\lambda^i(s) :=$ the smallest value of $\lambda$ for which both active and passive modes are equally desirable when the state is $s$. The index rule is then to order, at each time $m$, the current indices $\lambda^i(Y^i_m), 1 \leq i \leq N$, in decreasing order and render active the processes corresponding to the $M$ lowest indices, breaking ties as per some pre-specified rule, and render passive the remaining $N-M$ processes.

\subsection{Comparison with \cite{c1}}
\label{SSC:comparison:with:c1}
As mentioned in the introduction, this work draws upon \cite{c1} for its methodology, but there are some major differences. The first, of course, is what was already underscored in the introduction, viz., that \cite{c1} does not deal with a network situation with interference constraints and with possibility of spatial reuse of spectrum, as we do here. Another major difference is that \cite{c1} also had another control variable, viz., the number of packets to be transmitted. This leads to a different kind of issues that we do not face here. On the other hand, the network scenario with interference constraints opens up a whole new slew of complications, so that classical Whittle index theory is not directly applicable and one has to fall back upon some heuristics in order to adapt it for our purposes.

Even in the mathematical analysis, there are some important differences. The most important is the fact that in \cite{c1}, the state is real valued, whereas here we have stayed with the discrete valued queuing formalism. Where this matters is in the proof of the existence of a threshold policy, which is a key step in establishing Whittle indexability. There are three key properties of the value function which facilitate this: monotonicity, convexity and the increasing differences property. Of these, monotonicity can  be proved here just as in \cite{c1} by a pathwise comparison argument. The increasing difference property follows from convexity, but it is the convexity which becomes an awkward issue in the discrete domain. Thus the Whittle indexability that we invoke from \cite{c1} is to be understood only in an approximate sense-- it holds for the continuum analog which well approximates a large discrete queuing system. The latter is true because one can in fact view the state space of non-negative integers  as being embedded in reals. Then in the proof of convexity, complications arise only near the zero state. So if this state is visited only infrequently, the results for continuous state space carry over modulo some small approximation error.

Another point of departure with \cite{c1} we have here is the finite buffer length assumption which was not there in \cite{c1}.  This is a benign change, as it does not affect the proof of Whittle indexability  (in the continuous state space). Like discrete state space, it allows us to be more faithful to the real system.

\section{Scheduling Algorithms Based on Whittle Indices}
\label{SC:algorithms}
\subsection{Definition of Whittle-like Indices For Our Problem}
\label{SSC:whittle:for:our:problem}
We now use a procedure similar to Whittle's procedure~\cite{Whittle}, which was reviewed in Section~\ref{SSC:whittle:index:background}, to define Whittle-like indices for the problem formulated in Section~\ref{SSC:model:prob:formulation}. 
In this case (see \eqref{EQ:objective:function} and \eqref{EQ:interference:constraints}), (\ref{together}) gets replaced by
\begin{equation}
\limsup_{n\uparrow\infty}\frac{1}{n}E\left[\sum_{m=0}^{n-1}\sum_i(\nu_m^i f^i(Z^i_m) + C^iX^i_m + \lambda^i\hspace{-1em}\sum_{j\in\N^*(i)}\hspace{-1em}\nu^j_m)\right]
\label{together2}
\end{equation}
leading to the individual problems
\begin{equation}
\limsup_{n\uparrow\infty}\frac{1}{n}E\left[\sum_{m=0}^{n-1}\nu_m^i f^i(Z^i_m) + C^iX^i_m + \Lambda^i\nu^i_m\right] \label{separate2}
\end{equation}
for each $i$, with $\Lambda^i := \sum_{j\in\N^*(i)}\lambda^j$. Treating $\Lambda^i$'s as a surrogate for Whittle tax\footnote{$\approx$ negative subsidy, since this is a minimization problem unlike~\cite{Whittle}} that is `given', the problems decouple into individual problems and one can employ Whittle's logic to define a Whittle-like index, for a given state $j$, as that value of $\Lambda^i$ for which the active and passive modes are equally desirable at state $j$. Attractive as this scheme may appear, it is not without problems. For one thing, there is a non-trivial loss of information in the sense that the map from $\{\lambda^i\}$ to $\{\Lambda^i\}$ may not be invertible. Consider, e.g., a graph with two nodes, say $1$ and $2$, connected by an edge. Then $\N^*(1) = \N^*(2) = \{1,2\}$. So $\Lambda^1 =  \Lambda^2 = \lambda^1 + \lambda^2$. Hence in this example, the map from $\{\lambda^i\}$ to $\{\Lambda^i\}$ is not  invertible.

In concrete terms, what one finds is that moving over to $\{\Lambda^i\}$ may effectively change the constraint set itself. In particular, consider the problem of minimizing the cost in \eqref{EQ:objective:function} subject to the following  constraint:
\begin{equation}
\label{EQ:clique:interference:constraints}
\sum_{j \in \V}\nu^j_n \leq 1.
\end{equation}
Note that \eqref{separate2} are precisely the individual control problems  obtained by using the standard Whittle's procedure~\cite{Whittle} for decoupling the above constrained cost minimization problem, with $\{\Lambda^i\}$ being the corresponding Whittle-like indices. Also, the constraint in \eqref{EQ:clique:interference:constraints} is the interference constraint that we get if we treat the network as a \emph{clique}. This fact motivates the approach for computing Whittle indices that we present in Section~\ref{SSSC:Whittle:index:computation:clique}. Section~\ref{SSSC:Whittle:index:computation:original} presents an alternative scheme that has an additional tweak to circumvent the need to consider the clique model.

\subsection{Whittle Index Based Algorithm for Activating Users}
\label{SSC:activation:algorithm}
In this section, we provide an algorithm for selecting an independent set of users to activate in a given time slot, assuming that the Whittle indices of all the users in the slot have been already computed. In Section~\ref{SSC:Whittle:index:computation}, we provide two different approaches for computing Whittle indices-- the first by treating the graph representing the network as a clique (Section~\ref{SSSC:Whittle:index:computation:clique}) and the second based on interference constraints derived from the original graph (Section~\ref{SSSC:Whittle:index:computation:original}).

Suppose the Whittle indices\footnote{We denote the Whittle index of user $i$ under the clique model (respectively, model based on interference constraints derived from the original graph) as $\Lambda^i(X^i_n)$ (respectively, $\lambda^i(X^i_n)$). However, in Section~\ref{SSC:activation:algorithm}, for ease of exposition, we denote the Whittle index of user $i$ as $\lambda^i(X^i_n)$ regardless of which model is used to compute the Whittle indices.} $\lambda^i(X^i_n)$ of all the users $i \in \V$ have been computed in a given time slot $n$. An independent set of users to activate in the slot is selected as follows.

First, all users with empty queues are declared passive. Then,  those users $i \in \V$ for which $\lambda^i(X^i_n) \leq \lambda^j(X^j_n) \ \forall j \in \N(i)$ are declared active (ties are broken according to some tie-breaking rule, \emph{e.g.}, the user with smaller identifier (ID) is declared active). Next, for every active user $i$, all users $j \in \N(i)$ are declared passive.
In the next step, all users $i \in \V$ for which $\lambda^i(X^i_n) \leq \lambda^j(X^j_n) \ \forall j \in \N(i)$ and which are not yet declared passive, are declared active, and their neighbors are declared passive if already not so. This process is repeated till all users have been declared either active or passive.

Note that the set of users that are declared active constitute an independent set; these users transmit in the slot. Furthermore, for implementing the procedure, at any point in time, a user \emph{only requires information that is available with its neighboring users} and therefore the procedure can be implemented in a \emph{distributed} manner.

\subsection{Computation of Whittle Index}
\label{SSC:Whittle:index:computation}
\subsubsection{Computation Based on Constraints Derived from the  Clique Model}
\label{SSSC:Whittle:index:computation:clique}
Recall from Section~\ref{SSC:model:prob:formulation} that we represent the wireless network using a graph $\G = (\V, \E)$, which in general is not a clique. However, we now present an approach for computing Whittle indices by treating the graph as a clique, i.e., by assuming that a given user in $\V$ interferes with \emph{every other user} in $\V$. After Whittle indices have been computed using this approach, an independent set of users to activate in the slot is selected using the algorithm provided in Section~\ref{SSC:activation:algorithm}. The motivation for treating the graph as a clique is provided in Section~\ref{SSC:whittle:for:our:problem}.

Recall the dynamic programming equation for an individual queue $i$~\cite{RF:puterman:mdp}:
\begin{eqnarray}
V^i(x^i) & = & C^i x^i + \min_{\nu^i \in \{0,1\}} \left[ \nu^i f^i (x^i \wedge \Psi^i) \right.  \nonumber \\
& & + \sum_k V^i \left( \left[ x^i - \nu^i (x^i \wedge \Psi^i) + k \right] \wedge M^i \right) \mu^i(k)  \nonumber \\
& & \left. + (1 - \nu^i) \Lambda^i \right] - \beta^i,
\label{EQ:mdp:equation:clique}
\end{eqnarray}
where $V^i(\cdot)$ is the value function and $\beta^i$ is the optimal value of the average cost problem. The Whittle index, $\Lambda^i(X^i_n)$, for state $X^i_n = x^i$ is calculated by the following iteration (explained in Section~\ref{SSSC:computational:scheme:justification}):  At step $m$, solve the linear system of equations in variables $V^i(\cdot), \beta^i$ given by:
\begin{eqnarray*}
V^i(y^i) &=&\sum_kV^i(\left[y^i - y^i\wedge\Psi^i + k\right]\wedge M^i)\mu^i(k) - \beta^i \\
&&+\ C^iy^i + f^i(y^i\wedge\Psi^i), \ \ y^i \geq x^i,\\
V^i(y^i) &=& C^iy^i + \Lambda_m^i + \sum_kV^i(\left[y^i + k\right]\wedge M^i)\mu^i(k) \\
& & - \beta^i, \ x^i > y^i \neq 0, \\
V^i(0) &=& 0,
\end{eqnarray*}
and then perform a single iterate of
\begin{eqnarray*}
\Lambda_{m+1}^i & = & \Lambda_m^i + \gamma \left[f^i(x^i\wedge\Psi^i) - \Lambda_m^i \right. \\
& & +  \sum_k\mu^i(k)(V^i(\left[x^i - x^i\wedge\Psi^i + k\right]\wedge M^i) \\
&& \left. -  V^i(\left[x^i + k\right] \wedge M^i))\right].
\end{eqnarray*}
Here $\gamma > 0$ is a small learning parameter. Note that $y^i - y^i\wedge\Psi^i = \max(y^i - \Psi^i, 0)$.

\subsubsection{Computation Based on Constraints Derived from the Original Graph}
\label{SSSC:Whittle:index:computation:original}
We now present an approach for computing Whittle indices based on interference constraints derived from the original graph $\G = (\V, \E)$ (i.e., not by treating it as a clique).
From \eqref{EQ:interference:constraints}, we get $\sum_{j \in \N^*(i)}\nu^j_n = 1 \ \forall i.$
Relax it to:
$$\lim_{N\uparrow\infty}\frac{1}{N}\sum_{j\in\N^*(i)}E\left[\sum_{m=0}^{N-1}\nu^j_m\right] = 1.$$
By \eqref{EQ:objective:function} and using the Lagrange relaxation, the unconstrained problem has running cost:
$$\sum_{i \in \V} \left(C^i x^i + \nu^if^i(z^i) + \lambda^i \sum_{\{j: \ i \ \in \ \N^*(j)\}}\nu^j\right).$$
The dynamic programming equation for an individual queue $i$ is given by~\cite{RF:puterman:mdp}:
\begin{eqnarray}
V^i(x^i) & = & C^i x^i + \min_{\nu^i \in \{0,1\}} \left[ \nu^i f^i (x^i \wedge \Psi^i) \right.  \nonumber \\
& & + \sum_k V^i \left( \left[ x^i - \nu^i (x^i \wedge \Psi^i) + k \right] \wedge M^i \right) \mu^i(k)  \nonumber \\
& & \left. + (1 - \nu^i) \lambda^i \sum_{\{j: \ i \ \in \ \N^*(j)\}}\nu^j \right] - \beta^i,
\label{EQ:mdp:equation:original}
\end{eqnarray}
where $V^i(\cdot)$ is the value function and $\beta^i$ is the optimal value of the average cost problem. At time slot $n$, for each $i \in \V$, the Whittle index, $\lambda^i(X^i_n)$,  for state $X^i_n = x^i$ is calculated by the following iteration (explained in Section~\ref{SSSC:computational:scheme:justification}): At step $m$, solve the following linear system of equations in variables $V^i(\cdot), \beta^i$ for each fixed choice of $\nu^j$ such that $i \ \in \ \N^*(j)$,
\begin{eqnarray*}
V^i(y^i) &=& \sum_k V^i(\left[y^i - y^i \wedge\Psi^i + k\right]\wedge M^i)\mu^i(k) - \beta^i \\
&& + \ C^i y^i + f^i(y^i \wedge \Psi^i), \ \ y^i \geq x^i, \\
V^i(y^i) &=& \lambda^i_m \sum_{\{j : \ i \ \in \ \N^*(j)\}}\nu^j \\
&& + \sum_kV^i(\left[y^i + k\right] \wedge M^i)\mu^i(k) \\
&& + \ C^i y^i - \beta^i, \ \ x^i > y^i \neq 0,  \\
V^i(0) &=& 0,
\end{eqnarray*}
and then perform a single iterate of
\begin{eqnarray*}
\lambda^i_{m+1} &=& \lambda^i_m + \gamma\Big[f^i(x^i \wedge \Psi^i) - \lambda^i_m \sum_{\{j : \ i \ \in \ \N^*(j)\}}\nu^j \\
&& + \ \sum_k\mu^i(k)(V^i(\left[x^i - x^i \wedge \Psi^i + k\right] \wedge M^i) \\
&& - V^i(\left[x^i + k\right] \wedge M^i))\Big]
\end{eqnarray*}

Here we take $\sum_{\{j : \ i \ \in \ \N^*(j)\}}\nu^j$ to be the size of the maximum independent set in the subgraph formed by $\N(i)$.

In both  cases (Sections~\ref{SSSC:Whittle:index:computation:clique} and~\ref{SSSC:Whittle:index:computation:original}), for computational simplicity, the above iteration is performed for sufficiently large number of $x^i$ and then interpolated.

\subsubsection{Explanation for Above Computational Schemes}
\label{SSSC:computational:scheme:justification}
We comment briefly on the theoretical underpinnings  of the foregoing, omitting details which are analogous to \cite{c1}. To justify a Whittle-like index, one needs to establish Whittle indexability, i.e., the fact that the set of passive states depends monotonically on $\lambda^i$ for each $i$. For the first case (Section~\ref{SSSC:Whittle:index:computation:clique}), the situation is exactly as in \cite{c1} and in the second case (Section~\ref{SSSC:Whittle:index:computation:original}), there is an $i$-dependent scaling of $\lambda$ that does not affect the argument. The computational scheme in each case is similar to that of \cite{c1}: Whittle index is defined in terms of the value of $\lambda^i$ that, for given $x$, renders two functions of $x$ and $\lambda^i$ equal. The iterative scheme iterates candidate index values incrementally in the direction that decreases the difference between the two quantities and can be analyzed exactly as in \cite{c1}.

Despite the above similarities, note that there are significant differences between the contributions of~\cite{c1} and this paper, which are explained in Section~\ref{SSC:comparison:with:c1}.

\section{Simulations}
\label{SC:simulations}
In this section, we evaluate the performances of the proposed Whittle index based algorithms and compare them with those of the well known Slotted ALOHA~\cite{c7}, Max-Weight Scheduling (MWS)~\cite{RF:tassiulas:stability} and Lyapunov Drift~\cite{RF:energy:optimal:neely} algorithms via simulations. The performance is evaluated in terms of two metrics-- average cost and average total number of packets dropped per time slot-- at all the users in the network. We briefly review the Slotted ALOHA,  Max-Weight Scheduling and Lyapunov Drift Strategies in Section~\ref{SSC:Slotted:ALOHA:and:Max-Weight:Scheduling:Strategies} and present our simulation model and results in Sections~\ref{SSC:Simulation:Model} and~\ref{SSC:Simulation:results}, respectively.

\subsection{Slotted ALOHA,  Max-Weight Scheduling and Lyapunov Drift Strategies}
\label{SSC:Slotted:ALOHA:and:Max-Weight:Scheduling:Strategies}

\subsubsection{Slotted ALOHA}
\label{SSSC:Slotted:ALOHA}
Slotted ALOHA is a widely used randomized medium access control protocol~\cite{c7}. Under this protocol, whenever a given user has packets to transmit in a given time slot, it transmits them with probability $p$ and does not transmit them with probability $1 - p$ in the time slot, where $p \in (0,1)$ is a parameter. If two or more neighbouring users transmit in a given slot, then a collision takes place; in this case, each user involved in the collision repeats the above protocol, i.e., transmits (respectively, does not transmit) with probability $p$ (respectively, $1-p$), in the next time slot.

\subsubsection{Max-Weight Scheduling}
\label{SSSC:Max-Weight:Scheduling}
The Max-Weight Scheduling algorithm has been used in the context of resource allocation in wireless networks~\cite{RF:neely:survey},~\cite{RF:tassiulas:stability},~\cite{Tassiulas}, scheduling in input-queued switches~\cite{RF:mckeown:input:queued:switches} and several other contexts, and has been analytically shown to result in a high throughput and stability region in several prior works. Under the Max-Weight Scheduling algorithm, in each time slot $n$, each user is assigned a weight equal to the length of the queue of packets waiting at the transmitter that need to be sent to the receiver of that user. The weight of an independent set of users is defined to be the sum of the weights of the users in the independent set. In each time slot $n$, the independent set with maximum weight is found and each user belonging to that independent set transmits. Note that the computation of a maximum weight independent set is an NP-complete problem~\cite{RF:kleinberg:algorithm}. Hence practical implementation of the Max-Weight Scheduling algorithm is computationally prohibitive. Nevertheless, we use the Max-Weight Scheduling algorithm as a benchmark for comparison with our scheme.

\subsubsection{Lyapunov Drift Algorithm}
\label{SSSC:lyapunov}
The Lyapunov drift algorithm has been used to design stochastic optimal control policies in the context of resource allocation and routing problems in wireless networks to achieve system stability and performance optimization simultaneously~\cite{RF:energy:optimal:neely}. Under the Lyapunov drift algorithm, in each time slot $n$, if a user $i$ transmits, then a penalty that is proportional to $P_n^i$, which is the energy expended by the user for packet transmissions in slot $n$, is imposed  on the user. Also, in time slot $n$, user $i$ is assigned  weight $W_n^i = X_n^i Z_n^i - \theta P_n^i$, where $\theta$ is a nonnegative control parameter and $X_n^i$ and $Z_n^i$ are as defined in Section~\ref{SSC:model:prob:formulation}. In each time slot $n$, the independent set with maximum weight is found and each user belonging to that independent set transmits.

\subsection{Simulation Model}
\label{SSC:Simulation:Model}
In our simulations, we consider the model described in Section~\ref{SC:model:prob:formulation:background} with $L = 20$ users and buffer capacity $M^i = 100$ for each user $i \in \V$. The location of each user is selected uniformly at random in a square of dimensions $1$ unit $\times$ $1$ unit. Two users are neighbors iff the distance between them is less than $d_{threshold}$, where $d_{threshold}$ is a parameter. Throughout the simulations, we use $d_{threshold} = 0.6$ units. Let $\Psi^i$ be the maximum number of packets that may be transmitted by user $i$ in a given time slot. For the Whittle index based algorithms, we consider two cases: (i) $\Psi^i = \infty$, and (ii) $\Psi^i$ is uniformly chosen to be some value between $1$ and $M^i/ 5$ for user $i$ independent of other users. We refer to cases (i) and (ii) as the ``unrestricted transmission'' and ``restricted transmission'' cases, respectively. Note that in case (i), a user that transmits in a time slot sends \emph{all} the packets in its queue. Throughout our simulations, under the Slotted ALOHA,  Max-Weight Scheduling and Lyapunov Drift algorithms, the value $\Psi^i$ for user $i$ is the same as that in the restricted case of the Whittle index based algorithms. We assume that the number of packets, $\xi_n^i$, that arrive at user $i$ in time slot $n$ is a Poisson random variable with mean $l^i$; also, unless otherwise mentioned, $l^i$ is selected uniformly at random to be a value between $1$ and $M^i/10$ for each $i$ independent of other users. Also, in the Lyapunov drift algorithm, we use $\theta = 200$.

\subsection{Simulation Results}
\label{SSC:Simulation:results}
Henceforth, we refer to the algorithms based on  computation of the Whittle indices as in Sections~\ref{SSSC:Whittle:index:computation:clique} and~\ref{SSSC:Whittle:index:computation:original} as the ``Clique Whittle Policy'' and ``Graphical Whittle Policy'', respectively.

\begin{figure}[!hbt]
    \centering
     \resizebox{0.7\columnwidth}{!}{\includegraphics{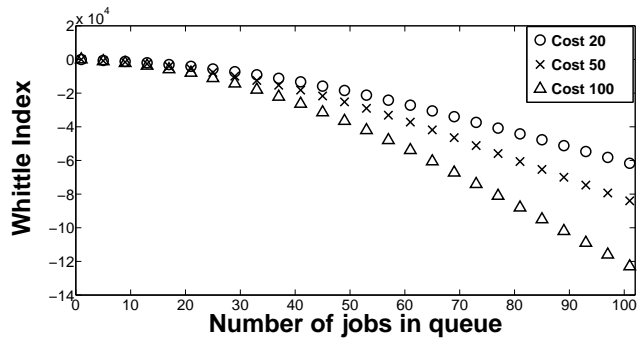}}
    \caption{Variation of Whittle Index versus number of jobs in the queue with holding costs 20, 50 and 100}
    \label{fig:Whittle:variation:holding:cost}
\end{figure}

Fig.~\ref{fig:Whittle:variation:holding:cost} shows the Whittle Index $\Lambda(x)$ for the Clique Whittle Policy versus the number of packets (jobs), $x$, in the queue for each of the holding cost values $C = 20, 50$ and $100$.  We see that for each value of $C$, $\Lambda(x)$ decreases in $x$. Also, when $C$ is increased, for a given value of $x$,  $\Lambda(x)$ decreases. Since under the algorithm described in Section~\ref{SSC:activation:algorithm}, we preferably select users with low values of $\Lambda(x)$ for transmission, the trends in Fig.~\ref{fig:Whittle:variation:holding:cost} show that users with high holding costs $C$ and large queue lengths $x$ are preferred for transmission. Intuitively, this leads to a low average cost, since the average cost is an increasing function of the holding cost $C$ as well as the queue length $x$ (see \eqref{EQ:objective:function}).
%



\begin{figure}[!hbt]
    \centering
     \resizebox{0.7\columnwidth}{!}{\includegraphics{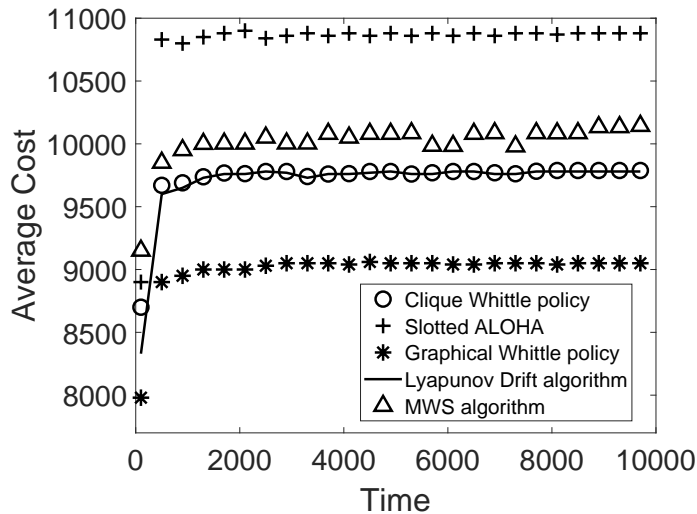}}
    \caption{Average cost under different algorithms for small arrival rates with restricted transmissions}
    \label{fig:Comparison:small:arrival:restricted1}
\end{figure}

\begin{figure}[!hbt]
    \centering
     \resizebox{0.7\columnwidth}{!}{\includegraphics{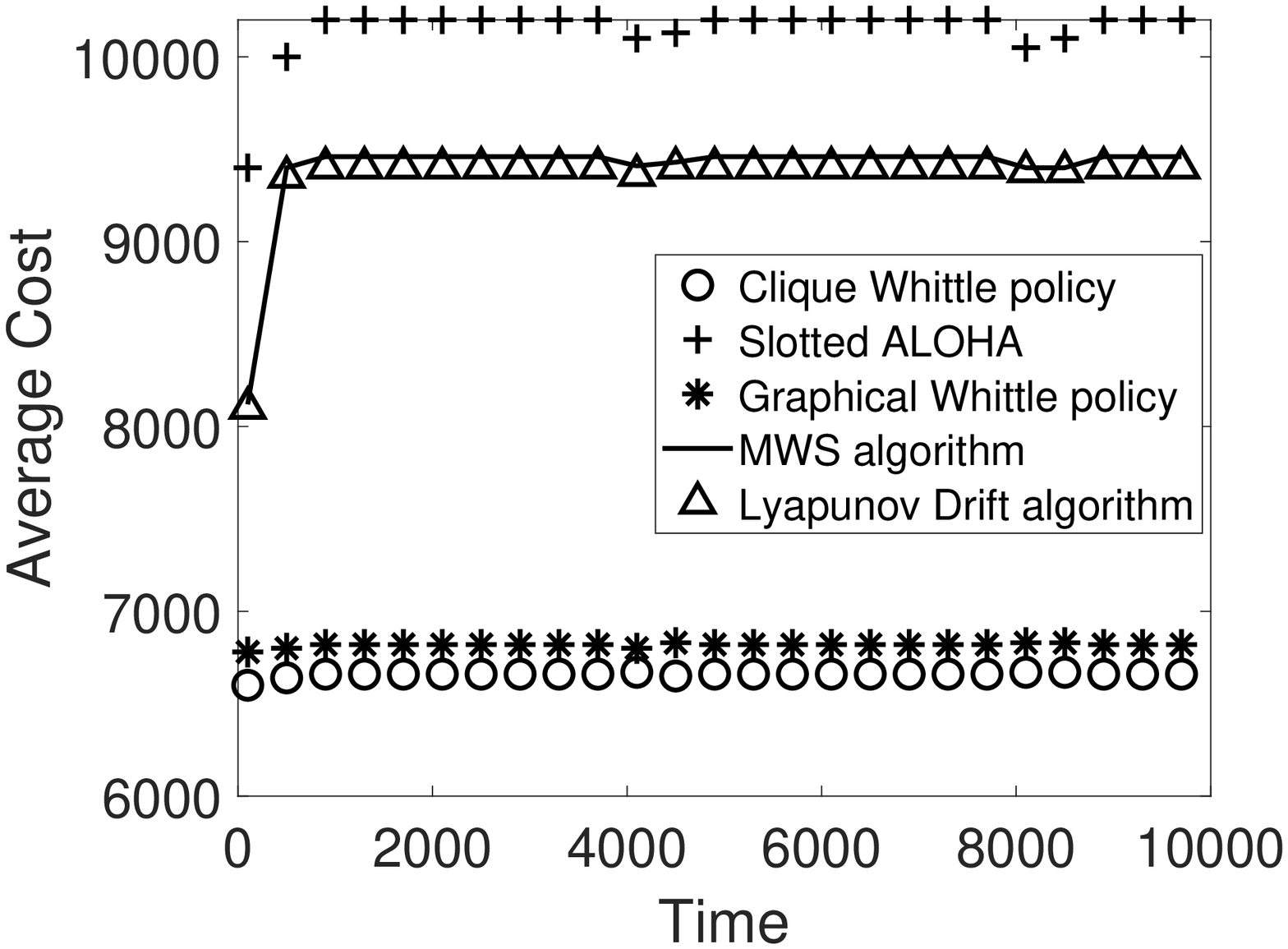}}
    \caption{Average cost under different algorithms for small arrival rates with unrestricted transmissions}
    \label{fig:Comparison:small:arrival:unrestricted1}
\end{figure}

\begin{figure}[!hbt]
    \centering
     \resizebox{0.7\columnwidth}{!}{\includegraphics{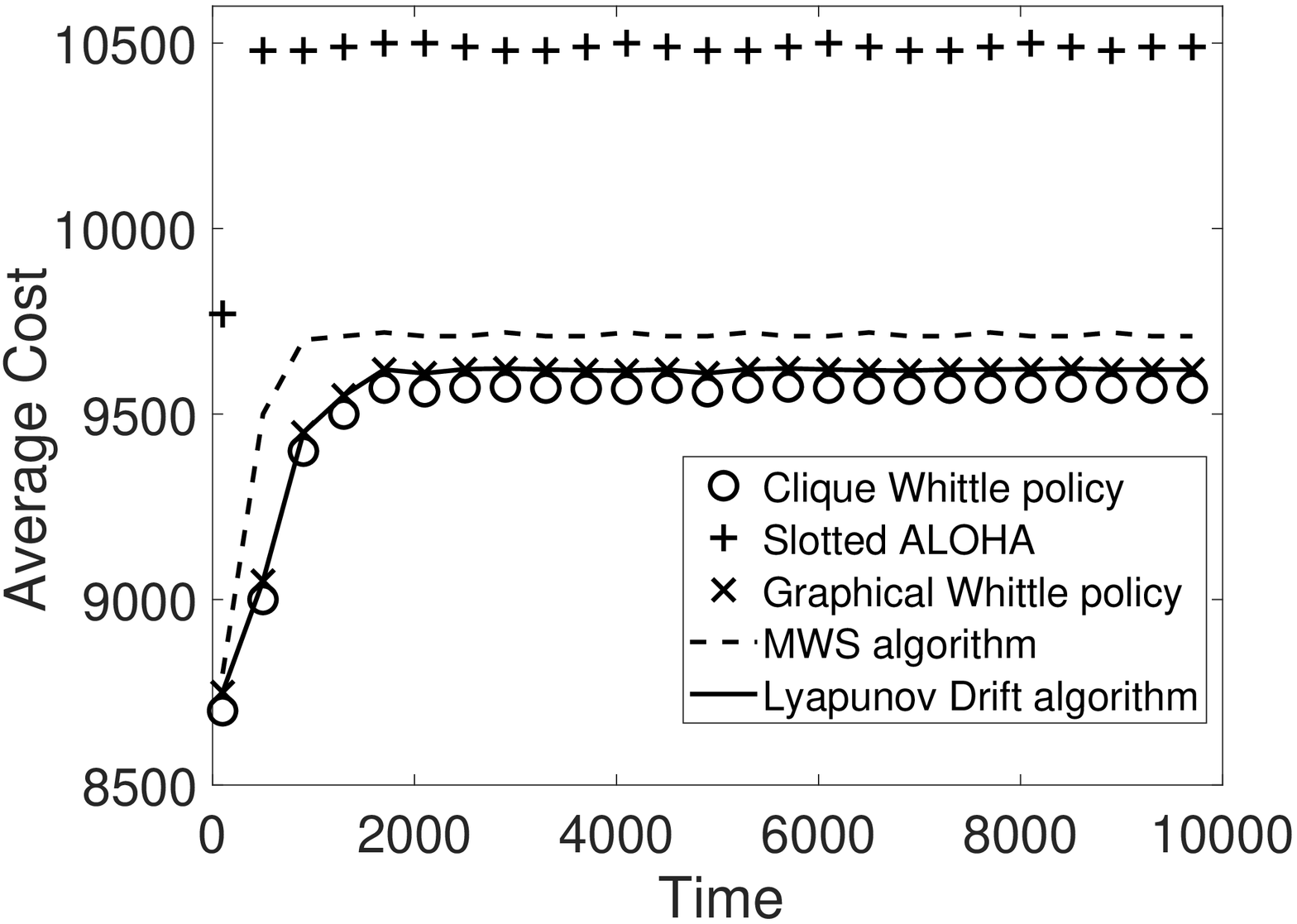}}
    \caption{Average cost under different algorithms for large arrival rates with restricted transmissions}
    \label{fig:Comparison:large:arrival:restricted1}
\end{figure}

\begin{figure}[!hbt]
    \centering
     \resizebox{0.7\columnwidth}{!}{\includegraphics{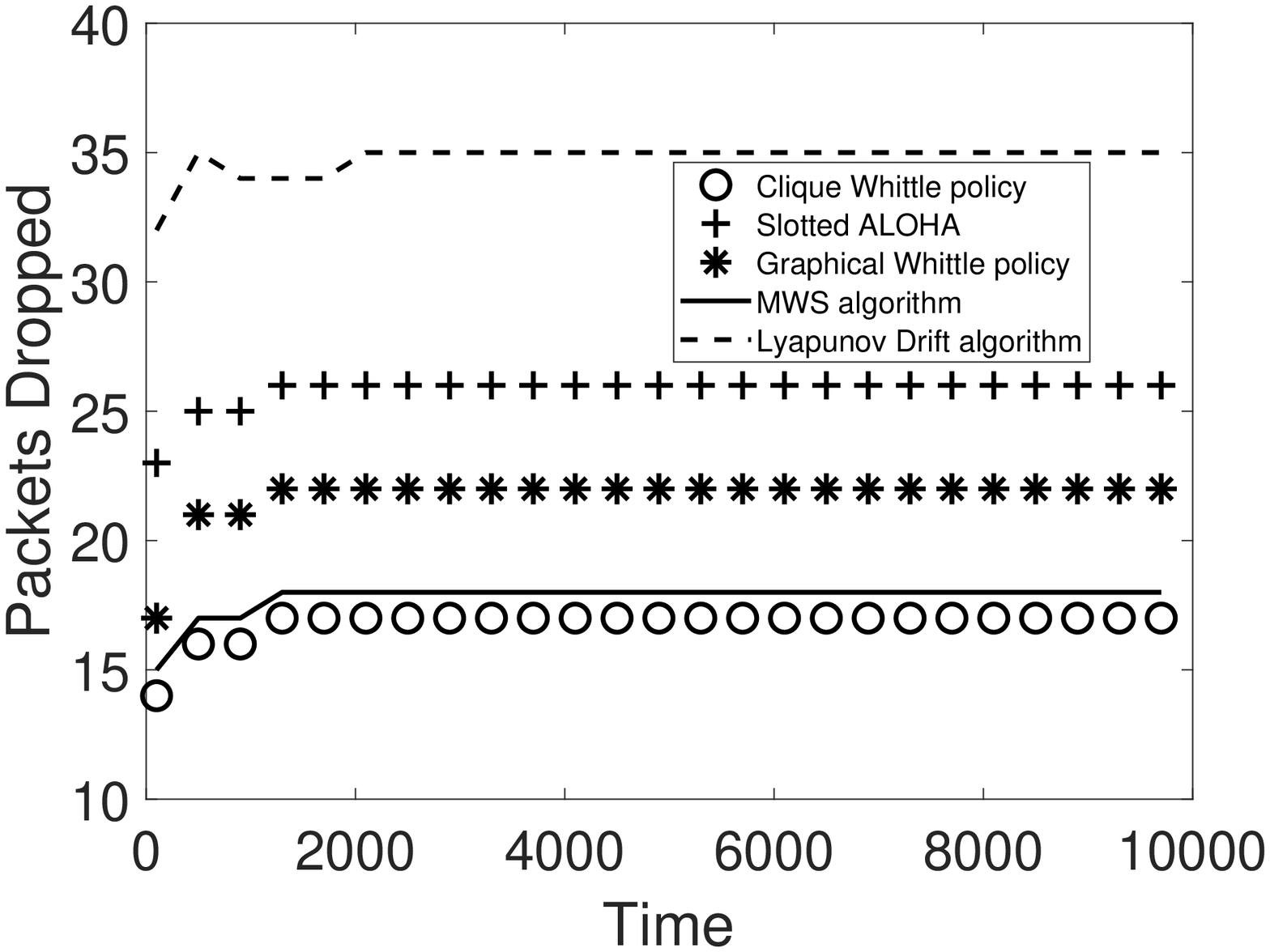}}
    \caption{Packets dropped under different algorithms for small arrival rates with restricted transmissions}
    \label{fig:Comparison:small:arrival:restricted2}
\end{figure}

\begin{figure}[!hbt]
    \centering
     \resizebox{0.7\columnwidth}{!}{\includegraphics{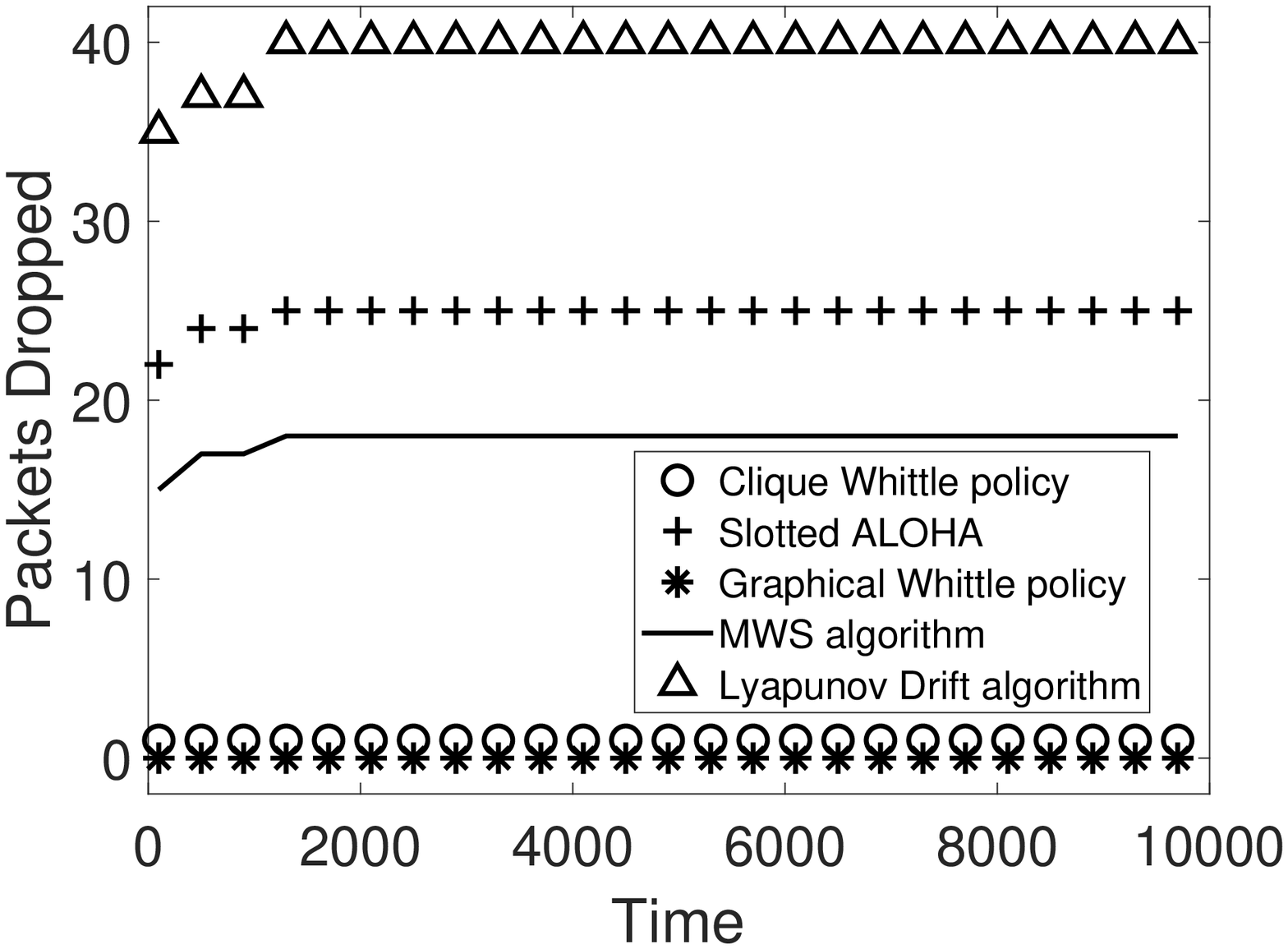}}
    \caption{Packets dropped under different algorithms for small arrival rates with unrestricted transmissions}
    \label{fig:Comparison:small:arrival:unrestricted2}
\end{figure}

\begin{figure}[!hbt]
    \centering
     \resizebox{0.7\columnwidth}{!}{\includegraphics{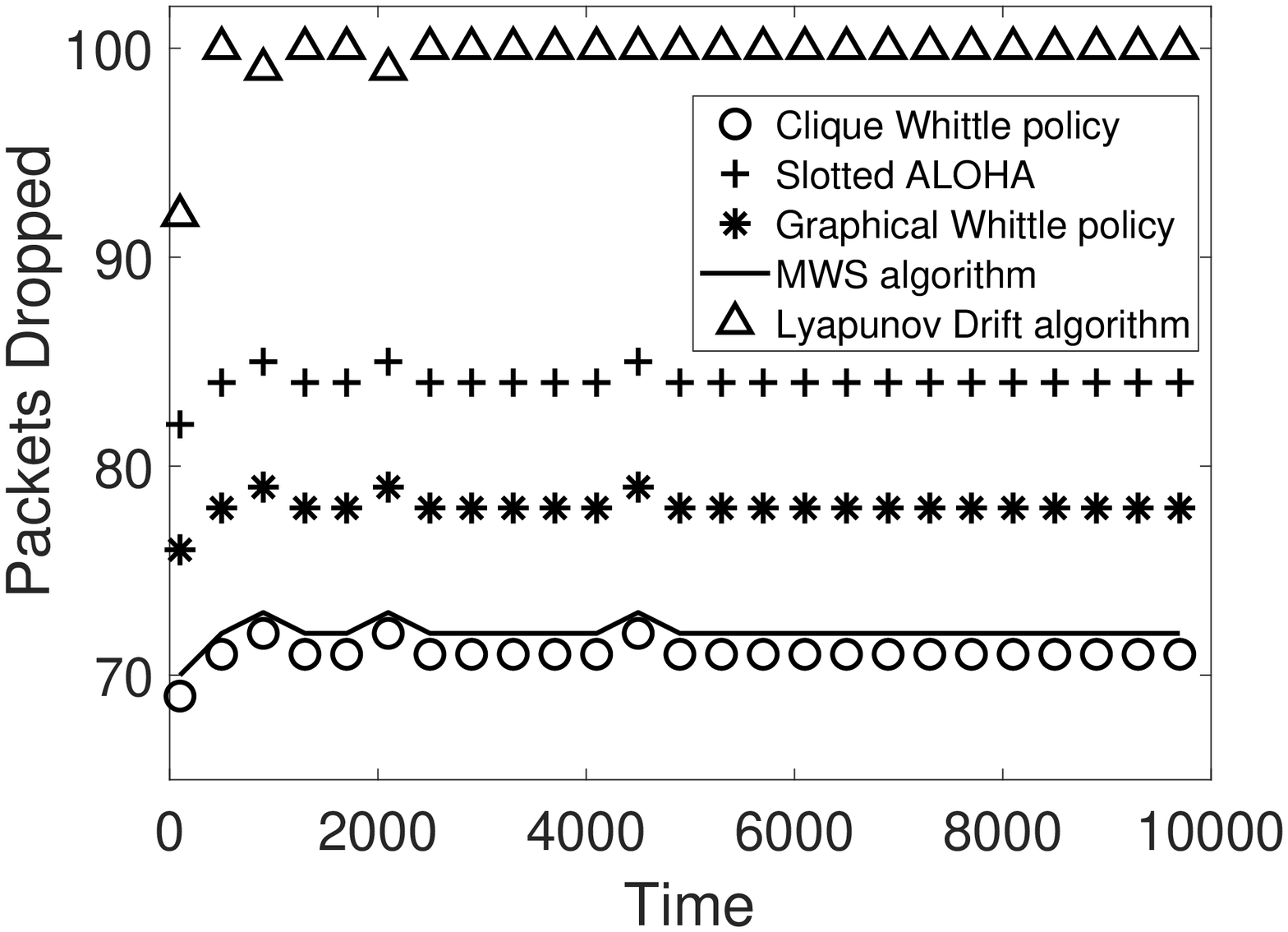}}
    \caption{Packets dropped under different algorithms for large arrival rates with restricted transmissions}
    \label{fig:Comparison:large:arrival:restricted2}
\end{figure}

Next, by ``large arrival rates'' (respectively, ``small arrival rates''), we mean  that $l^i$ is chosen uniformly at random between $1$ and $M^i/6$ (respectively, between $1$ and $M^i/15$).
Figs.~\ref{fig:Comparison:small:arrival:restricted1},~\ref{fig:Comparison:small:arrival:unrestricted1} and~\ref{fig:Comparison:large:arrival:restricted1} (respectively,~\ref{fig:Comparison:small:arrival:restricted2},~\ref{fig:Comparison:small:arrival:unrestricted2} and~\ref{fig:Comparison:large:arrival:restricted2}) compare the performances of the Clique Whittle Policy, Graphical Whittle Policy, Slotted ALOHA algorithm,  Max Weight Scheduling (MWS) algorithm and Lyapunov Drift algorithm in terms of average cost (respectively, average total number of packets dropped at all the users in the network per time slot)  for the cases with (i) small arrival rates and restricted transmissions, (ii) small arrival rates and unrestricted transmissions, and (iii) large arrival rates and restricted transmissions, respectively. Fig.~\ref{fig:Comparison:small:arrival:restricted1} (respectively,~\ref{fig:Comparison:small:arrival:restricted2}) shows that in the case with small arrival rates and restricted transmissions, the Graphical Whittle Policy (respectively, Clique Whittle Policy) performs the best in terms of average cost (respectively, packets dropped).
Figs.~\ref{fig:Comparison:small:arrival:unrestricted1} and~\ref{fig:Comparison:small:arrival:unrestricted2} show that in the case with small arrival rates and unrestricted packet transmissions, both the Whittle index based policies significantly outperform the Slotted ALOHA algorithm, the MWS algorithm and the Lyapunov drift algorithm in terms of both average cost and packets dropped. Also, in this case, the packets dropped under the two Whittle index based policies are close to zero, which is consistent with intuition, since $\Psi^i = \infty$, due to which users transmit all the packets in their queue whenever they transmit in a slot.
Figs.~\ref{fig:Comparison:large:arrival:restricted1} and~\ref{fig:Comparison:large:arrival:restricted2} show that in the case of large arrival rates with restricted transmissions, the Clique Whittle Policy outperforms the other four algorithms in terms of average cost and packets dropped, respectively. Overall, Figs.~\ref{fig:Comparison:small:arrival:restricted1}-\ref{fig:Comparison:large:arrival:restricted2} show that \emph{both the Whittle index
based policies outperform the Slotted ALOHA algorithm in all the cases and
both the MWS and the Lyapunov drift algorithms in most of the cases considered}. Transmissions in practice will always be restricted and although the performance of our proposed policies in the unrestricted transmissions case was superior to the other three schemes for low arrival rates, it was not so for high arrival rates. This is presumably because the  errors due to ad hoc tweaks in both variants of our policies become more pronounced in the high traffic (i.e., high arrival and transmission rates) regime. 



 \begin{figure}[!hbt]
    \centering
     \resizebox{0.7\columnwidth}{!}{\includegraphics{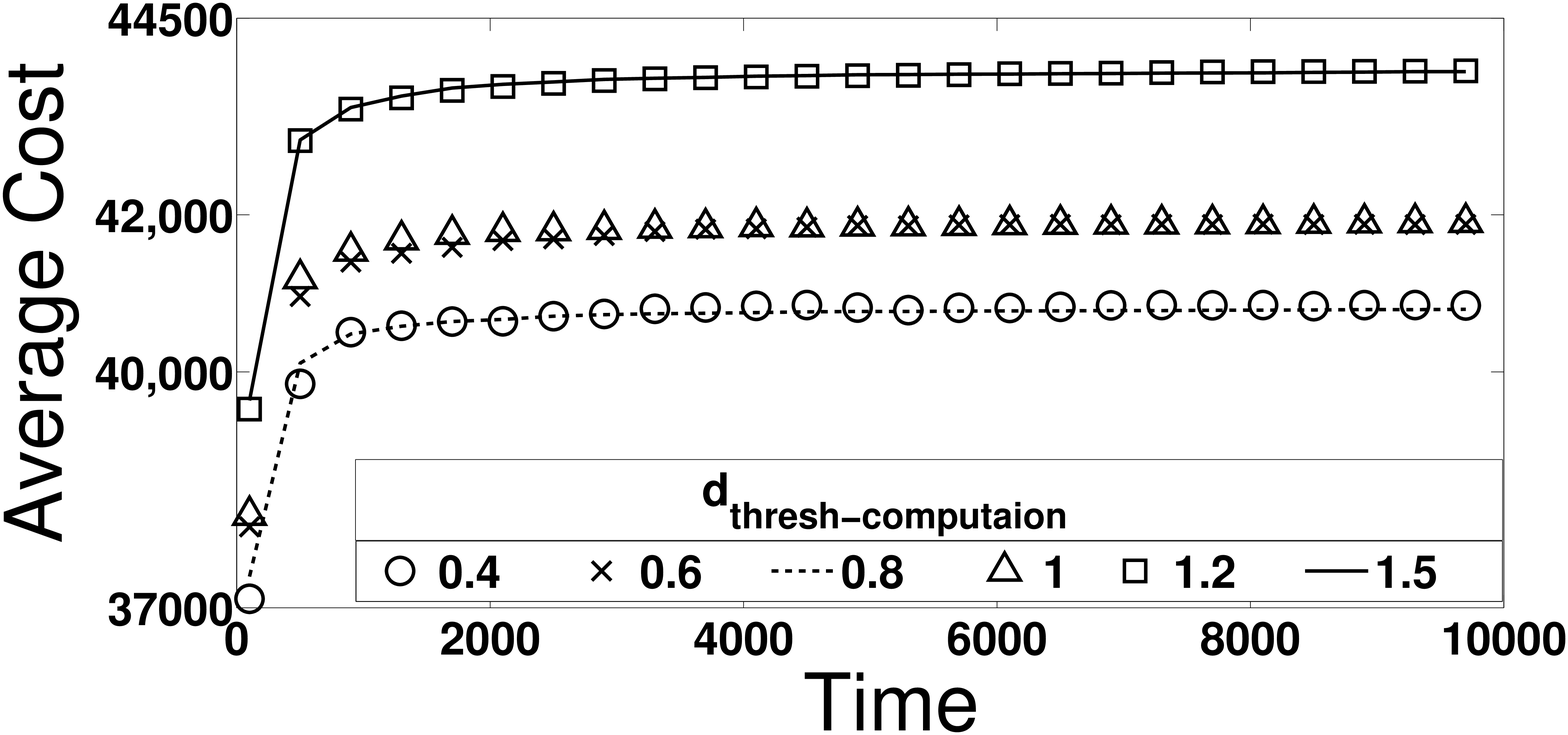}}
    \caption{Average cost for the generalized Whittle policy with different values of the parameter $d_{thresh-computation}$ for restricted transmissions case}
    \label{combined:limited1}
\end{figure}
 \begin{figure}[!hbt]
    \centering
     \resizebox{0.7\columnwidth}{!}{\includegraphics{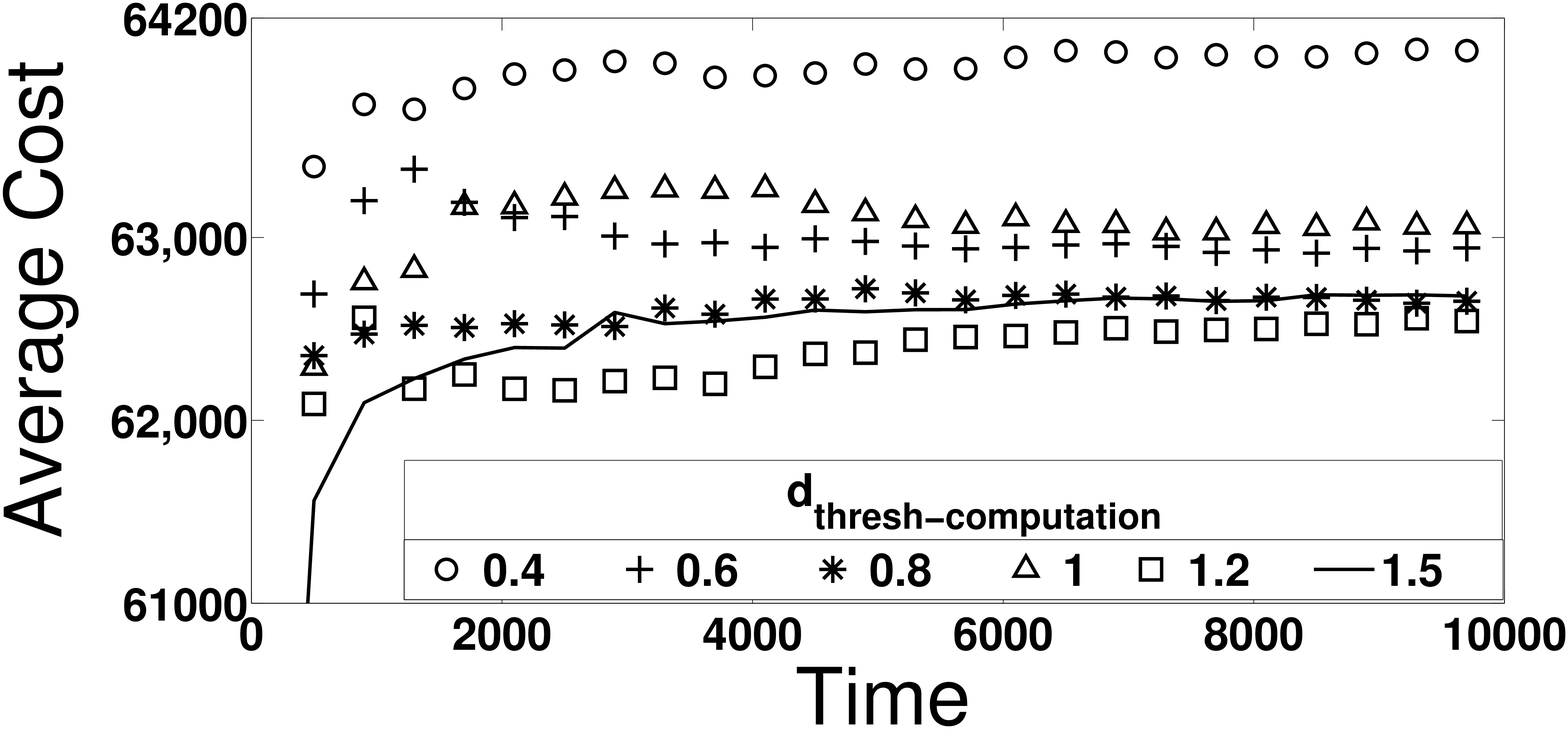}}
    \caption{Average cost for the generalized Whittle policy with different values of the parameter $d_{thresh-computation}$ for unrestricted transmissions case}
    \label{combined:unlimited1}
\end{figure}

 \begin{figure}[!hbt]
    \centering
     \resizebox{0.7\columnwidth}{!}{\includegraphics{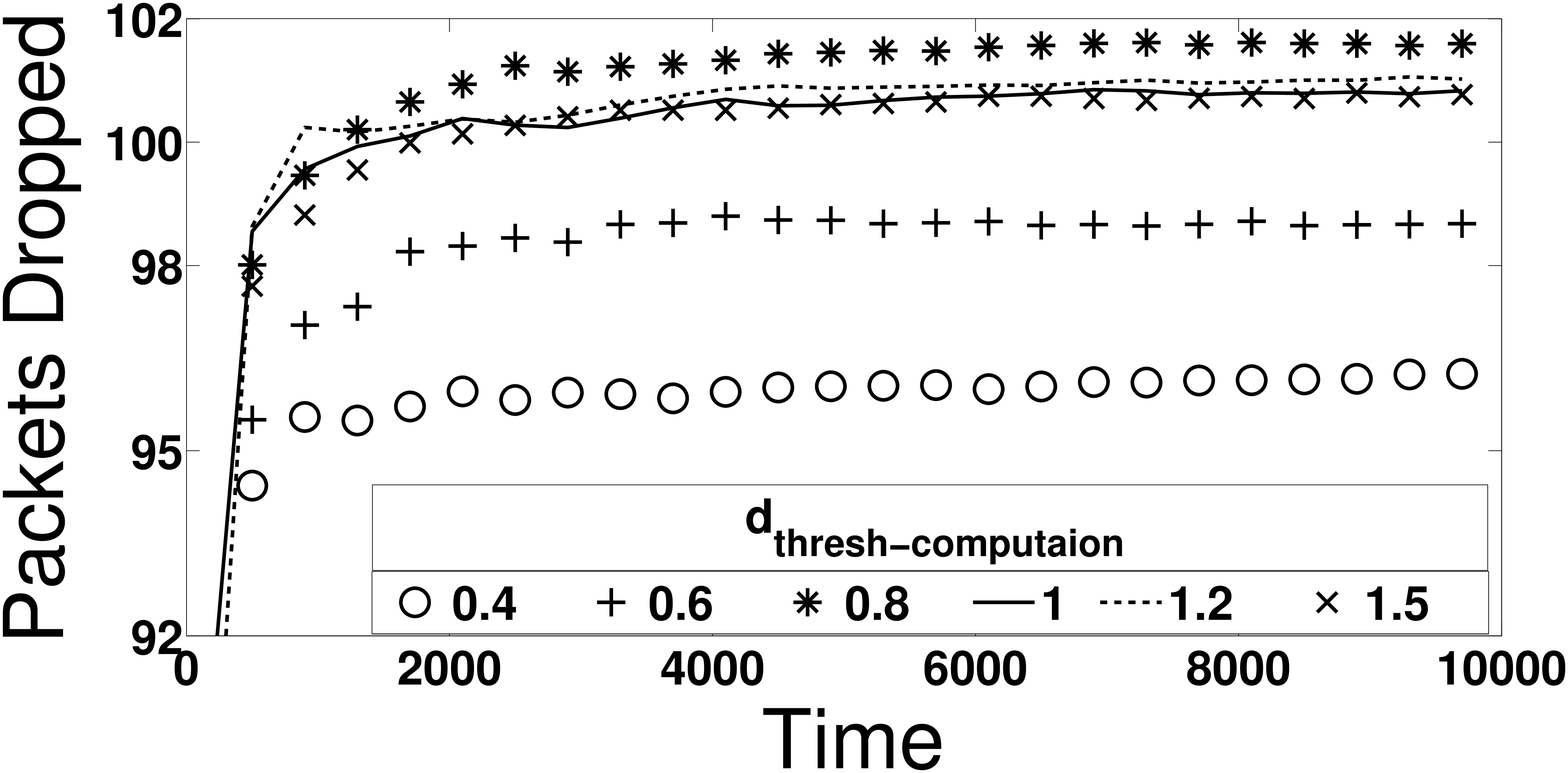}}
    \caption{Packets dropped for the generalized Whittle policy with different values of the parameter $d_{thresh-computation}$ for restricted transmissions case}
    \label{combined:limited2}
\end{figure}
 \begin{figure}[!hbt]
    \centering
     \resizebox{0.7\columnwidth}{!}{\includegraphics{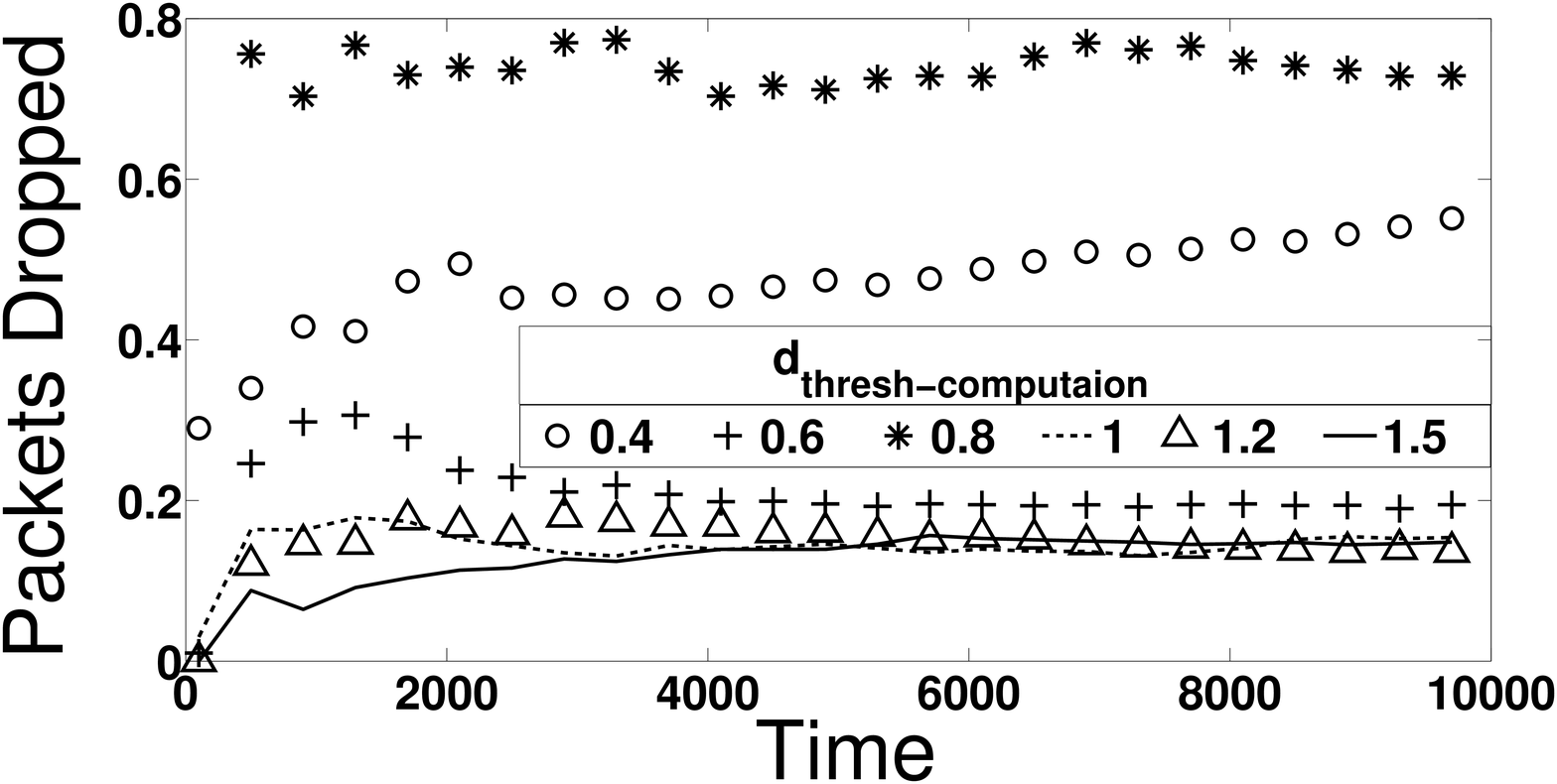}}
    \caption{Packets dropped for the generalized Whittle policy with different values of the parameter $d_{thresh-computation}$ for unrestricted transmissions case}
    \label{combined:unlimited2}
\end{figure}

%
%

Finally, we consider the following common generalization of the Clique Whittle and Graphical Whittle policies: \emph{during the computation of the Whittle indices}, instead of using the constraints derived from the original graph as in Section~\ref{SSSC:Whittle:index:computation:original}, constraints derived from the graph in which two users are neighbors iff the distance between them is less than a parameter $d_{thresh-computation}$ are used. However, \emph{when users actually transmit}, as before, the transmissions of two users interfere iff the distance between them is less than $d_{threshold}$. Note that since all users are located in a $1$ unit $\times$ $1$ unit square, the special case $d_{thresh-computation} = d_{threshold}$ (respectively, $d_{thresh-computation} = 1.5$) of the above generalized policy corresponds to  the Graphical Whittle policy (respectively, Clique Whittle policy). We investigate as to what values of $d_{thresh-computation}$ result in the best performance. For the parameter values  $\Psi^i = 20$ and $C^i = 20$ $\ \forall i \in \V$, Figs.~\ref{combined:limited1} and~\ref{combined:unlimited1} (respectively,~\ref{combined:limited2} and~\ref{combined:unlimited2}) show the average cost (respectively, average total number of packets dropped at all the users in the network per time slot) under the policies with different values of $d_{thresh-computation}$ in the restricted transmissions and unrestricted transmissions cases, respectively.
In the restricted transmissions case, the average cost as well as packets dropped are minimized when $d_{thresh-computation} = 0.4$. In the unrestricted transmissions case, the average cost as well as packets dropped are minimized when $d_{thresh-computation} = 1.2$. Thus, by using plots such as those in Figs.~\ref{combined:limited1}-\ref{combined:unlimited2}, we can find out as to what values of $d_{thresh-computation}$ result in the best performance for given parameter values.

\section{Conclusions and Future Work}
\label{SC:conclusions}
We proposed two Whittle index based scheduling policies for  scheduling packet transmissions with the objective of minimizing the energy consumption and data transmission delay of users in a wireless network in which spatial reuse of spectrum is employed. The first policy treats the graph as a clique and the second policy is based on interference constraints derived from the original graph. We evaluated the performance of these two policies via extensive simulations, in terms of average cost and packets dropped, and showed that they outperform the well-known Slotted ALOHA, maximum-weight scheduling and  Lyapunov drift algorithms. A direction for future research is to use techniques similar to those developed in this paper to design Whittle index based policies for other resource allocation problems in wireless networks wherein spatial reuse of spectrum is employed.

\bibliographystyle{IEEEtran}
\bibliography{ref}

\end{document}